\documentclass[twocolumn, 10pt, DIV=15]{scrartcl}
\usepackage{textcomp}
\usepackage[english]{babel}
\usepackage[utf8]{inputenc}
\usepackage[T1]{fontenc}
\usepackage{csquotes} 
\usepackage{ragged2e}
\usepackage{microtype}
\usepackage{lmodern}
\usepackage{lipsum}
\usepackage[marginal,norule]{footmisc}
\usepackage{mathtools} 

\usepackage{amsfonts}
\usepackage{amssymb}
\usepackage{siunitx}
\usepackage{braket}
\usepackage{nicefrac}
\usepackage{graphicx}
\graphicspath{{./Pictures/}}
\usepackage{placeins}
\usepackage{xcolor}
\usepackage{wallpaper} 
\usepackage{xpatch}
\usepackage{array}
\usepackage{booktabs}

\usepackage{wrapfig}
\usepackage{caption}
\usepackage{subcaption}
\usepackage{floatrow}
\usepackage[
  style = numeric-comp,
  sorting = none,
  doi = false,
  url = false,
  isbn = false,
  bibencoding = utf8,
  backend = biber,
]{biblatex}
\usepackage{times}
\usepackage[pdfusetitle]{hyperref}
\title{\large Moiré lattice of twisted bilayer graphene as template for non-covalent functionalization}

\author{Tobias Dierke\(^{\ast,1}\), Stefan Wolff\(^{\ast,1}\), Roland Gillen\(^{1}\), Jasmin Eisenkolb\(^{2}\), Tamara Nagel\(^{2}\), Sabine Maier\(^{1}\),\\ Milan Kivala\(^{3}\), Frank Hauke\(^{2}\), Andreas Hirsch\(^{2}\) and Janina Maultzsch\(^{1}\)\\
}
\date{\footnotesize{\(^{1}\)\textit{Department of Physics, Chair of Experimental Physics, Friedrich-Alexander-Universität Erlangen-Nürnberg, Staudtstraße 7, 91058 Erlangen, Germany}\\
\(^{2}\)\textit{Department of Chemistry and Pharmacy, Chair of Organic Chemistry II \& Center of Advanced Materials and Processes, Friedrich-Alexander-Universität Erlangen-Nürnberg, Nikolaus-Fiebiger-Str. 10, 91058 Erlangen, Germany}\\
\(^{3}\)\textit{Organisch-Chemisches Institut, Universität Heidelberg, Im Neuenheimer Feld 270, 69120 Heidelberg, Germany}} \\tobias.dierke@fau.de\\stefan.wolff@fau.de\\
\today}

\makeatletter

\renewcommand\sectionlinesformat[4]{%
  \@hangfrom{\hskip #2#3}{\MakeUppercase{#4}}%
}
\makeatother

\setkomafont{disposition}{\rmfamily\bfseries\centering}
\setkomafont{section}{\small}
\setkomafont{title}{\normalfont\rmfamily\bfseries}
\setkomafont{author}{\normalfont\rmfamily\normalsize}

\usepackage[autooneside=false]{scrlayer-scrpage}
  \setkomafont{pagehead}{\scshape}
  \setkomafont{pagefoot}{\itshape\small}
    \clearpairofpagestyles
  \automark[subsection]{section}
  \makeatletter
  \lohead{}
  \cohead{}
  \rohead{\pagemark}
  \rofoot{\copyright{} 2024}
  \cofoot{}
  \makeatother
  \recalctypearea

\begin{document}
\maketitle
\begin{abstract}
We present a novel approach to achieve spatial variations in the degree of non-covalent functionalization of twisted bilayer graphene (tBLG). The tBLG with twist angles varying between \(\sim\)\,\(5^\circ\) and \(7^\circ\) was non-covalently functionalized with 1,4,5,8,9,11-hexaazatriphenylenehexacarbonitrile (HATCN) molecules. Our results show a correlation between the degree of functionalization and the twist angle of tBLG. This correlation was determined through Raman spectroscopy, where areas with larger twist angles exhibited a lower HATCN peak intensity compared to areas with smaller twist angles. We suggest that the HATCN adsorption follows the moiré pattern of tBLG by avoiding AA-stacked areas and attach predominantly to areas with a local AB-stacking order of tBLG, forming an overall ABA-stacking configuration. This is supported by density functional theory (DFT) calculations. Our work highlights the role of the moiré lattice in controlling the non-covalent functionalization of tBLG. Our approach can be generalized for designing nanoscale patterns on two-dimensional (2D) materials using moiré structures as a template.
\end{abstract}

\section{Introduction}
The ability to manipulate graphene by chemical functionalization has opened up a new field of research, including structuring of graphene areas into spatial patterns.
A variety of approaches can be used to pattern the basal plane of graphene.
For instance, spatial patterning of covalently functionalized graphene can be achieved by the use of plasma jets, both with and without the use of masks
\cite{Ye2014,Hernndez2013}, by generating reactive radicals through laser irradiation \cite{Bao2021,Edelthalhammer2020}, or by patterning the underlying substrate on which graphene is deposited \cite{Dierke2022}.
However, the challenge lies in controlling the spatial attachment of the functional groups on a nanometer scale \cite{Edelthalhammer2020,Dierke2022}, which is a key requirement for exploiting the full potential of graphene-based devices.
On the other hand, well-defined nanoscale patterns were successfully achieved by self-assembly of carbon-based molecules via non-covalent interactions on graphene by nanomanipulation in ambient scanning tunneling microscopy (STM) \cite{Tahara2018,delaRie2022}.
Such non-covalent interactions via \(\pi\)-\(\pi\) stacking are becoming a focus of recent research.

Twisted bilayer graphene (tBLG) with its moiré superlattice exhibits periodic changes in the local stacking order and electronic band structures, leading to remarkable phenomena such as flat bands \cite{Bistritzer2011}, correlated insulating states \cite{Cao2018,Xie2020}, and superconductivity \cite{Lu2019,Yankowitz2019}, thus invoking much research in the field of twisted bilayers, twistronics and moiré structures.
At narrow angles between \(4^\circ\) and \(10^\circ\), even small variations of the twist angle result in large changes in the size of the moiré superlattice. At even smaller angles of \(\sim1^\circ\), self-organized reconstruction at an atomic scale into periodic stacking domains may occur \cite{Yoo2019}.
A recent idea is to use the moiré pattern of twisted structures as templates for designing elaborate patterns at the nanoscale {\cite{Hidalgo2021,Majumdar2023}. For example, the adsorption energies of non-planar methylamine and methanethiol molecules and molecular intercalations on tBLG have been shown to depend on the twist angle \cite{Hidalgo2019,Araki2022}.

In this work, we present a novel approach for achieving spatial variations in the degree of non-covalent functionalization of tBLG by using the moiré lattice as a template for \(\pi\)-\(\pi\) stacking. The exfoliated twisted bilayer graphene was non-covalently functionalized with 1,4,5,8,9,11-hexaazatriphenylenehexacarbonitrile (HATCN) molecules.
HATCN has a planar structure with an aromatic center and six outer cyano groups. 
Adopting a planar adsorption configuration on graphene, HATCN molecules can only interact with each other through a weak dipolar coupling between their partially negative charged nitrogen atoms and the positively charged carbon atoms in the neighboring cyano groups \cite{Mller2019,Christodoulou2014,Jeong2016,Oh2017}.
Therefore, it is expected that the van-der-Waals (vdW) and \(\pi\)-\(\pi\) interaction with graphene can be stronger for a planar adsorption.
The twist angle and degree of non-covalent functionalization of the tBLG sample are analyzed by Raman spectroscopy. Our results show a correlation of the twist angle and the degree of HATCN functionalization. Thus, the moiré lattice plays a predominant role for the adsorption of the HATCN molecules.
Density functional theory (DFT) calculations suggest preferred stacking configurations of HATCN molecules with tBLG, supporting our interpretation of the moiré lattice as template for non-covalent functionalization.

\section{Results and Discussion}

\begin{figure}
     \centering
     \includegraphics[width=\linewidth]{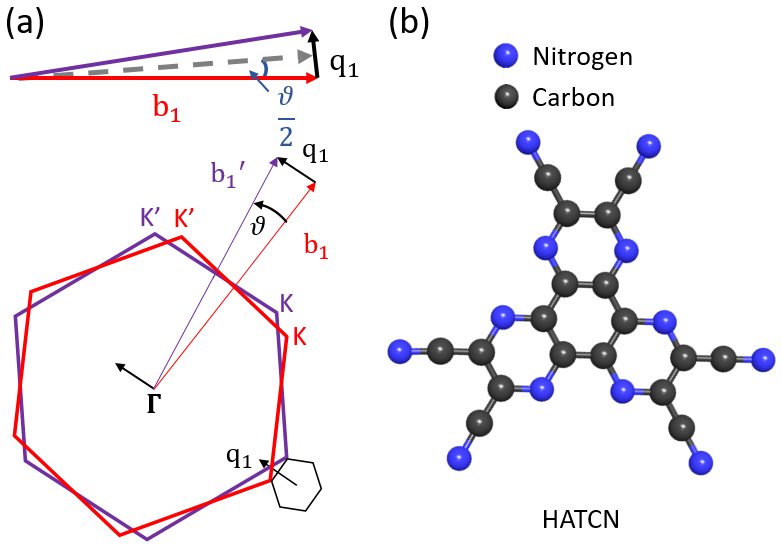}
     \caption{(a) Brillouin zone of twisted bilayer graphene with corresponding mini Brillouin zone and rotational wavevector \(\mathbf{q_1}\). (b) Molecular structure of 1,4,5,8,9,11-hexaazatriphenylenehexacarbonitrile (HATCN).}
     \label{fig:Bild5}
\end{figure}

As a first step, we analyze the twist angle of the entire tBLG flake via Raman mapping of the so-called \(R^{\prime}\) mode. Figure~\ref{fig:Bild5}\,(a) shows the Brillouin zones of rotated bilayer graphene, with the reciprocal lattice vectors of the bottom and top graphene \(\mathbf{b_1}\) and \(\mathbf{b^{\prime}_1}\), respectively. Their difference is the reciprocal rotation lattice vector \(\mathbf{q_1}\), which defines the `mini' Brillouin zone, see Figure~\ref{fig:Bild5}\,(a). For small twist angles \(\vartheta\), the direction of \(\mathbf{q_1}\) can be approximated as perpendicular to \(\mathbf{b_1}\) or \(\mathbf{b^{\prime}_1}\), thus pointing to the \(\mathit{\Gamma K}\) direction of the individual layers' Brillouin zone. This rotational wavevector \(\mathbf{q_1}\) of the twisted Brillouin zone allows additional Raman scattering away from the \(\mathit{\Gamma}\) point. As an example, the branches of the phonon dispersion intersecting the green line in Figure~\ref{fig:Bild1}\,(a) correspond to \(\mathit{\Gamma}\)-point modes in the tBLG and can now be measured in the Raman spectrum. The frequencies of these modes depend on the twist angle; they are often referred to as moiré phonons. The absolute value of the rotational wavevector \(\mathbf{q_1}\) depends on the twist angle [see Figure~\ref{fig:Bild5}\,(a)]:
\begin{equation}
\sin\biggl(\frac{\vartheta}{2}\biggr)=\frac{\nicefrac{q_1}{2}}{b_1}
\end{equation}
with \(b_1=|\mathbf{b_1}|=\frac{4\pi}{\sqrt{3}a}\), with \(a=2.46\)\,\AA{} lattice constant of graphene, resulting in 
\begin{equation}
q_1(\vartheta)=\frac{8\pi}{\sqrt{3}a}\sin\biggl(\frac{\vartheta}{2}\biggr).
\label{eq:winkel}
\end{equation}
Thus, the phonon wavevector along the \(\mathit{\Gamma K}\) direction can be replaced by the twist angle. For larger twist angles (\(\vartheta>10^{\circ}\)) the rotational wavevector \(\mathbf{q_1}\) points more and more towards the \(\mathit{\Gamma M}\) direction, i.e. the conversion of \(\vartheta\) to a resulting phonon frequency has to be adjusted for certain branches in the phonon dispersion \cite{Carozo2011}. In this work, tBLG areas with twist angles of around 6\(^\circ\) are investigated. Therefore, we use here the graphene phonon dispersion in \(\mathit{\Gamma K}\) direction to derive the \(\vartheta\)-dependent frequencies of the moiré phonons. The \(R^{\prime}\) mode (\(\sim 1620\,\text{cm}^{-1}\)), which originates from the longitudinal optical (LO) phonon branch, is the most distinct twist-induced mode in the Raman spectrum, due to the strong electron-phonon coupling for the LO branch near the \(\mathit{\Gamma}\)-point \cite{Carozo2011}, see Figure~\ref{fig:Bild1}\,(c). For the determination of the twist angle in our samples, Raman maps of the \(R^{\prime}\) frequency are analyzed and converted to a twist-angle map, based on the phonon dispersion in Figure~\ref{fig:Bild1}\,(b).

\begin{figure}
     \centering
     \includegraphics[width=\linewidth]{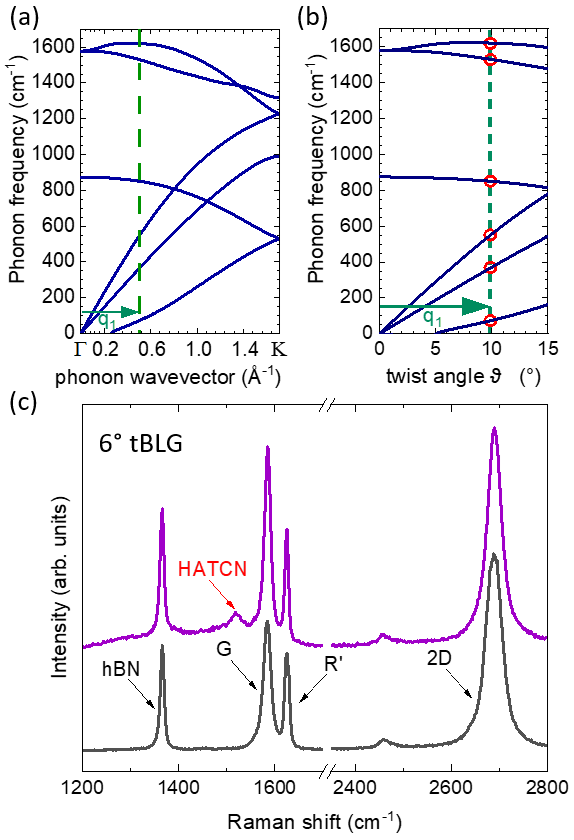}
     \caption{(a) Calculated phonon dispersion of single-layer graphene. The green arrow indicates the rotational wavevector \(\mathbf{q_1}\). (b) Phonon dispersion as a function of twist angle \(\vartheta\). Red circles indicate the moiré phonons for a given \(\mathbf{q_1}\) [same as in (a)].(c) Raman spectra (\(\lambda=532.17\)\,nm) of a \(\sim6^\circ\) twisted bilayer graphene sample with corresponding \(R^{\prime}\) mode, before (grey) and after (purple) HATCN functionalization. The characteristic mode of the HATCN molecule is indicated.}
     \label{fig:Bild1}
\end{figure}

\ref{fig:Bild2}\,(a) shows the Raman map of the \(\mathit{2D}\)-mode frequency, indicating the layer number of different areas in our sample.
This agrees with the optical microscope image in SI Figure S1.
The area framed by the blue line is the twisted bilayer area. The corresponding \(R^{\prime}\)-mode position and resulting twist-angle map is shown in Figure~\ref{fig:Bild2}\,(b), see also SI Figure S2.
We observe twist-angle variations in our structure between \(\sim5.3^\circ\) and \(\sim6.7^\circ\).
Red arrows point to areas with a twist angle of \(\sim6.3^\circ\) compared to the other parts in the twisted bilayer area with a slightly smaller twist angle of \(\sim5.5^\circ\).
Similar variations of the twist angle within a given tBLG structure have been reported in \cite{Schpers2022}.
The green line indicates an area (white) of the tBLG flake where no \(R^{\prime}\) mode was observed. We attribute this to a twist angle below \(\lesssim 5.3^\circ\).

To investigate the degree of strain in our tBLG sample, we analyze the peak position of the Raman \(G\) mode of graphene, which shows a rather homogeneous distribution across the twisted bilayer area (see SI Figure S3). From this we exclude that the variations in the \(R^{\prime}\)-mode frequency [\ref{fig:Bild2}\,(b)] are due to strain \cite{Mohiuddin2009}.
If the shift of the \(R^{\prime}\) peak of \(\sim4\)\,cm\(^{-1}\)  was caused by strain, this would correspond to about \(\sim0.4\)\,\% strain variations in the twisted bilayer area, assuming the same strain-dependence as for the graphene \(G\) mode. Because of the uniform distribution of \(G\)-mode frequencies across the entire tBLG area, we exclude such strain variations.

\begin{figure*}
     \centering
     \includegraphics[width=\linewidth]{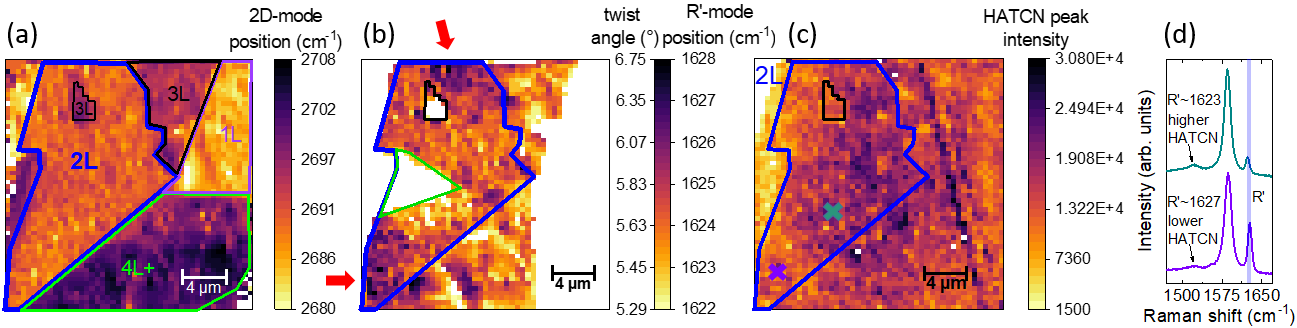}
     \caption{(a) Raman map of the \(\mathit{2D}\)-mode frequency after non-covalent functionalization. The \(\mathit{2D}\) mode was fitted by a single Lorentzian for all different graphene layers. Areas with different layer numbers are marked in the map. The area framed by the blue line shows the twisted bilayer (2L) area. (b) Raman map of the \(R^{\prime}\)-mode frequency and corresponding twist angle after non-covalent functionalization. The green framed area shows no \(R^{\prime}\) mode because the twist angle is smaller than \(\lesssim5.3^\circ\). Red arrows point to areas with higher \(R^{\prime}\)-mode frequency and therefore larger twist angle. There is no significant change in the twist angle map before and after the chemical treatment.
     (c) Raman map showing the intensity of the HATCN peak. The green and purple crosses mark the position of the spectra in Figure~\ref{fig:Bild2}\,(d). Exemplary Raman spectra of tBLG with non-covalent HATCN functionalization, highlighting the intensity difference of the HATCN mode at varying \(R^{\prime}\)-mode positions.}
     \label{fig:Bild2}
\end{figure*}

\ref{fig:Bild1}\,(c) shows the Raman spectra from the same tBLG flake before (grey) and after (purple) functionalization with HATCN.
After functionalization, there is no increase in the defect-induced \(D\) mode visible, as expected for a non-covalent functionalization. Instead, a Raman mode belonging to HATCN molecules appears at \(\sim 1520\,\text{cm}^{-1}\) (see SI Figure S20). In order to determine the relative amount of HATCN molecules on the graphene surface, the intensity of the HATCN mode is analyzed in Figure~\ref{fig:Bild2}\,(c). 
We observe a clear correlation between the HATCN-mode intensity and the twist angle: 
Areas with larger twist angle show a lower HATCN-mode intensity compared to areas with smaller twist angle. Figure~\ref{fig:Bild2}\,(d) shows exemplary spectra from an area with twist angle \(\sim\) 6.3 \(^\circ\) (\(R^{\prime}\) mode at \(\sim1627\,\)cm\(^{-1}\), higher intensity of the HATCN mode) and twist angle \(\sim\) 5.4\(^\circ\) (\(R^{\prime}\) mode at \(\sim1623\,\)cm\(^{-1}\), lower intensity of the HATCN mode).
The correlation between the twist angle and the intensity of the HATCN mode can also be observed in other regions of the sample, such as the 3L and 4L+ area, which contain the twisted graphene layer.
Additionally, in Figure~\ref{fig:Bild2}\,(c) the HATCN-mode intensity is pronounced along a line across the sample, which can be attributed to a fold in the hBN layer, see the optical microscope image in SI Figure S1.
Here we observe also a high intensity of the HATCN signal. This might be attributed to a weaker attachment of the graphene to the hBN next to the fold, see also SI Figure S4.

The correlation between twist angle and the HATCN-mode intensity points to a twist-angle dependent arrangement of the HATCN molecules on the tBLG.
We expect a low coverage of HATCN molecules, leading to a `planar' monolayer adsorption on the graphene surface (also due to hBN, see SI Figure S16), indicating a supressed molecule-molecule interaction.\cite{Christodoulou2014}
In Figure~\ref{fig:Bild4} we show the geometries of the moiré lattices for twist angles of \(5^\circ\) and \(7^\circ\), the upper and lower boundaries of twist angles in our sample.

For the same area, the number of AA-stacked and AB-stacked regions at \(7^\circ\) (right) is higher than at \(5^\circ\) twist angle (left). The moiré superlattice for \(5^\circ\) (\(7^\circ\)) has a periodicity of 2.82\,nm (2.02\,nm). The size of the moiré unit cell for \(7^\circ\) coincides in fact with the unit cell of the self-assembly of HATCN on Ag(111).\cite{Mller2019} Assuming that the non-covalent functionalization with HATCN molecules will arrange itself following the size and local stacking arrangements of the moiré lattice, would initially suggest that the HATCN molecules form this self-assembly on the \(7^\circ\) structure, thus giving the highest amount of funtionalization for this configuration. However, the measurements show the highest amount of funtionalization for a twist angle of \(5^\circ\), see Figure~\ref{fig:Bild2}.

An investigation of the total energy (SI Figure S7) confirms that the HATCN molecules seem to avoid AA-stacked regions and prefer stacking on AB-stacked bilayer graphene regions, see DFT calculations below. This is consistent with the electronic charge density difference of twisted graphene, which varies spatially.\cite{Hidalgo2021}
Our DFT calculations of the charge density redistribution of HATCN on bilayer graphene (see SI Figures S5 and S6) show the highest interaction strength of HATCN molecules which are placed on AB-stacked bilayer graphene to form an ABA configuration.
Additionally, the energy difference per HATCN between the optimal ABA configuration and a less suitable ABB arrangement is on the order of 0.1\,eV. This makes a significant change of the molecular arrangement, including a transition to the self-assembly configuration, unlikely.

Figure~\ref{fig:Bild4} schematically shows a possible configuration of HATCN molecules with an AB-stacking of the HATCN centering on AB-stacked tBGL domains. The distance between the nitrogen atoms at the outer wings of the molecules remains in the order of several \AA{} for the \(5^\circ\) twist angle configuration. At \(7^\circ\), the distance between the nitrogen atoms of neighboring HATCN molecules decreases to around 1-2\,\AA. As a result the HATCN molecules cannot attach to every AB-stacked region on the \(7^\circ\) moiré lattice since the partially negatively charged nitrogen atoms of the CN groups are too close to each other.

The fact that the experimental HATCN-mode intensity is higher at \(5^\circ\) than at \(7^\circ\), indicates that the surface-molecule interaction dominates over the molecule-molecule interaction, which also aligns with our DFT calculations. The hypothesis of preferential attachment of HATCN molecules on AB-stacked regions in tBLG is in agreement with the HATCN molecule distribution shown in Figure~\ref{fig:Bild2}\,(c). This clearly indicates that the moiré lattice of tBLG can affect the non-covalent interaction of molecules on graphene.

To further test our hypothesis, we performed DFT calculations of \(6 \times 6\) bilayer graphene supercells in which a single HATCN molecule was placed. The starting arrangement of bilayer graphene was chosen to be either AA- or AB-stacked. The HATCN molecule was placed on top of the bilayer graphene such that the center of the molecule aligns with the upper graphene layer in either AA or AB configuration, resulting in overall AAA-, AAB-, ABB-, and ABA-stacking configurations, see SI Figure S9.

After relaxation of the structures, an investigation of the total energy reveals that the energetically most favorable arrangement exhibits an ABA-stacking order, which arises from the two configurations in which the HATCN molecule is placed in an AB arrangement on the top layer of both AA- and AB-stacked bilayer graphene. Note that throughout the relaxation process of the AA-stacked bilayer graphene with an HATCN molecule placed in an AB configuration, the interaction between the molecule and the graphene is strong enough to induce a transition in the bilayer graphene from an AA- to an AB-like configuration (see SI Figure S6). This results in a total energy comparable to that of the ABA-stacking arrangement, which exhibits the lowest total energy.

The highest total energy is found for the AAA-stacking configuration (see SI Figure S7). In contrast to the supercell described above where the HATCN molecule is in AB configuration on top of AA-stacked graphene, no major structural changes occur during the relaxation process. This hints at a lower interaction strength between the molecule and the AA-stacked graphene.

In addition to the \(6\times 6\) supercells, we also constructed a \(5\times 5\) and a \(7\times 7\) supercell, each containing one HATCN molecule (all in ABA configuration). The relaxation of these structures gives further insight into the closest possible arrangement of neighboring HATCN molecules. The \(5\times 5\) cell seems to be insufficient for the accommodation of a HATCN molecule, which leads to a partial detachment of the molecule (see SI Figure S10). This is furthermore reflected by the fact that the energy per HATCN is even higher than the energy per HATCN for the unfortunate AAA configuration, see SI Figure S7. The energy per HATCN of the \(7\times 7\) structure matches the energy of the corresponding \(6\times 6\) configuration, hinting at the fact that larger distances between the molecules do not result in significantly less favorable arrangements.

The interaction between HATCN and tBLG is confirmed by calculating the redistribution of the charge density of individual HATCN molecules on bilayer graphene. The charge densities of isolated bilayer graphene and isolated HATCN molecules were subtracted from the charge density of the entire structure. A much more additional charge density between graphene and HATCN is found for the ABA-configuration compered to other arrangements (see SI Figure S5).

Our DFT calculations of freestanding HATCN confirm the self-assembly, however with a unit cell of 1.88\,nm (see SI Figure S11). The distances between neighboring CN-groups remain larger than 3.5\,\AA. The atoms of the HATCN molecules were artificially kept within the \(xy\)-plane, since a relaxation would otherwise not have been possible. This furthermore illustrates the high importance of the interaction of HATCN with the substrate.

Our DFT results clearly indicate stronger interactions for an AB-alignment of the HATCN molecule on top of graphene, which suggests that this is the more favorable arrangement compared to AA-stacking. Additionally, because of the significantly lower total energy of the ABA configurations, this seems to be the most favorable one. These results hint at the preferred formation of an ABA arrangement of tBLG and HATCN molecules, in agreement with the experimental data.

\begin{figure*}
     \centering
     \includegraphics[width=\linewidth]{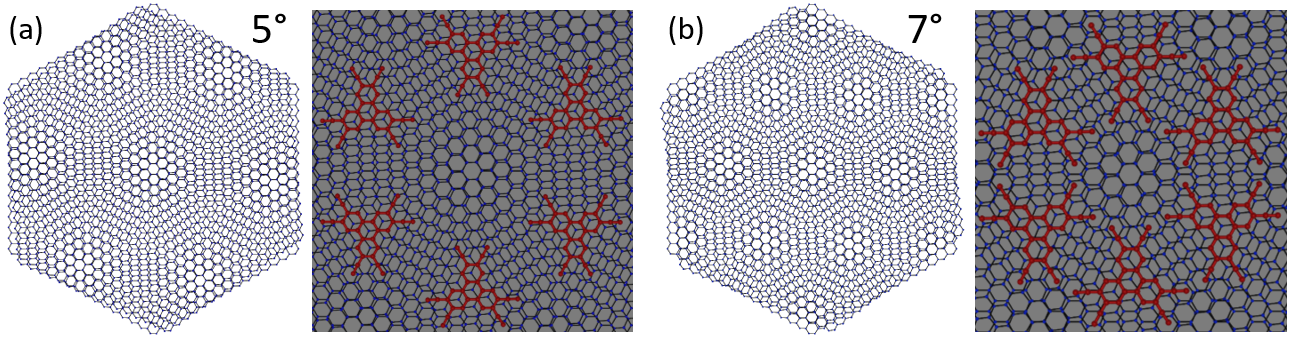}
     \caption{(a) and (b); moiré lattice of  \(5^\circ\) and \(7^\circ\) twisted bilayer graphene, respectively. Schematic views of possible arrangements of HATCN molecules (red) on AB-stacked regions, showing different densities of the HATCN molecules for the two twist angles.
     In the schematic view of the \(7^\circ\) moiré lattice, the nitrogen atoms of the CN groups are too close to each other. As a result, the HATCN molecules cannot attach to each AB-stacked region. For additional arrangements see SI Figure S8.}
     \label{fig:Bild4}
\end{figure*}

To obtain more insight into the stability of the HATCN molecule adsorption, we performed temperature-dependent Raman measurements at elevated temperatures.
The Raman maps of the twist angle distribution (see SI Figs. S12-S17) show no significant difference at temperatures up to 200\(^\circ\)C compared to room temperature [\ref{fig:Bild2}\,(b)].
\ref{fig:Temp50-100} shows a comparison of the HATCN-mode intensity maps at 50\(^\circ\)C (a) and 100\(^\circ\)C (b).   
At higher temperatures, the HATCN molecules become more mobile and are able to move on the graphene surface; in Figure~\ref{fig:Temp50-100}\,(b), the intensity of the HATCN signal changes locally. 
For example, the degree of HATCN functionalization increases locally in the area marked by the green circle. In the other parts, the distribution is still similar to the 50\(^\circ\)C and the room-temperature Raman map. 
The area showing now higher HATCN signal, also exhibits a slightly smaller twist angle, leading to larger distances between the HATCN molecules and therefore to a preferred attachment. 
Thus, the HATCN molecules seem to still arrange themselves through the influence of the underlying moiré lattice and  
the correlation between twist angle; the degree of the HATCN functionalization is still visible in the Raman maps at 100\(^\circ\)C. 
At higher temperatures, a homogeneous distribution of HATCN molecules in the tBLG region is observed (SI Figure S16).
Further temperature-dependent Raman maps, as well as maps showing the twist angle and the HATCN-mode intensity after a second HATCN functionalization are shown in the SI.

The question might arise whether the HATCN molecules are intercalated between the graphene layers or between the bottom hBN and tBLG. We therefore prepared functionalized graphene with an additional top hBN layer (after functionalization). Temperature-dependent Raman measurements (SI Figure S18) show that the HATCN mode is still present at 300\(^\circ\)C, in contrast to our previous measurements without the top hBN layer. This indicates that the HATCN molecules do not intercalate but are indeed on top of the tBLG structure.

 \begin{figure}
     \centering
     \includegraphics[width=\linewidth]{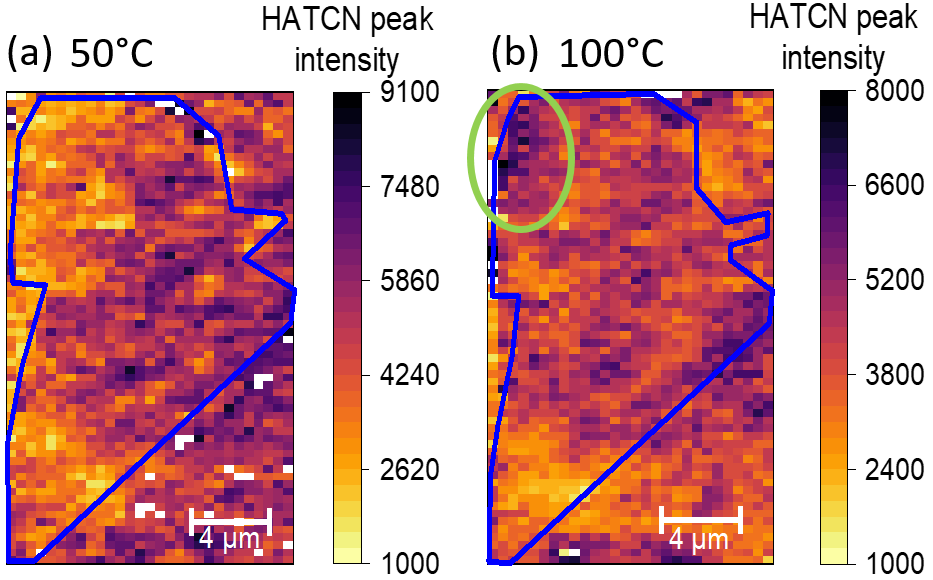}
     \caption{Comparsion of the HATCN peak intensity distribution at 50\(^\circ\)C (left) and 100\(^\circ\)C (right) of the same tBLG sample shown in Figure~\ref{fig:Bild2}. A clear increase of the HATCN peak intensity is observed in the upper left part of the twisted bilayer area (green circle) at 100\(^\circ\)C.}
     \label{fig:Temp50-100}
 \end{figure}

In order to rule out an accidential inhomogeneity of the HATCN-molecule distribution on our tBLG samples, we performed the same non-covalent functionalization procedure on single-layer and natural bilayer graphene. 
These measurements show a homogeneous intensity distribution of the HATCN Raman signal, indicating a homogeneous coverage of HATCN molecules (see SI Figure S19). For bilayer graphene, we observe an additional influence of the underlying substrate:
The total intensity of the HATCN mode is lower for graphene on hBN than on \(\mathrm{SiO_2}\) (see SI Figure S4). This is due to the chemical inertness of hBN and is in line with our previous investigations \cite{Dierke2022,Schfer2016} on covalently functionalized graphene, showing that the reactivity of the underlying substrate plays an important role also for the degree of the non-covalent functionalization.

\section{Conclusion}

In summary, we present a novel approach to achieve spatial variations in the degree of non-covalent functionalization of bilayer graphene by using the moiré lattice of tBLG as a template.
Our results show a clear correlation between the twist angle of tBLG and the degree of functionalization with HATCN molecules. 
Our assumption of a preferred attachment of the HATCN molecules in AB-stacked regions, resulting in a local ABA-stacking with the twisted graphene layers, is supported by DFT calculations.
Moiré structures of other 2D materials may be used as well for nanoscale patterning of non-covalent functionalization. 
The ability to precisely control the spatial attachment of functional groups by tuning the twist angle in twisted bilayer materials provides new possibilities in the field of high-precision nanoscale materials engineering.

\section{Samples and Methods}

We prepared twisted bilayer graphene on hexagonal boron nitride (hBN) via mechanical exfoliation using a stacking method similar to the one described in \cite{Kim2016}: hBN and graphene flakes are exfoliated onto silicon wafers with 300\,nm \(\mathrm{SiO_2}\). The exfoliated graphene flake is then cut into two parts by a laser cutter.
One part is picked up by a polycarbonate (PC)/polydimethylsiloxane (PDMS) stamp and held in a fixed position. The Si wafer with the second graphene part is rotated by the desired angle. This rotated second part is picked up by the stamp holding the first part of the graphene flake, thereby creating a twisted bilayer with the twist angle defined by the rotation. As an underlying substrate, we have chosen hBN, which is picked up in the last step. Unlike \(\mathrm{Si/SiO_2}\) substrates, it lacks dangling bonds, does not exhibit charge puddles and is atomically flat. Due to the flatness and a small lattice mismatch between hBN and graphene, the graphene experiences minor strain variations compared to graphene on \(\mathrm{SiO_2}\)\cite{Neumann2015,Dean2010,Mayorov2011}. 

The twisted graphene - hBN heterostructure is then non-covalently functionalized with 1,4,5,8,9,11-hexaazatriphenylenehexacarbonitrile (HATCN) - electron acceptor molecules, see Figure~\ref{fig:Bild5}\,(b). HATCN is dissolved in acetonitrile (\(10^{-4}\)\,mol/L); the twisted bilayer sample is immersed in the solution for seven days and afterwards rinsed with isopropyl alcohol (IPA). 
The HATCN was synthesized according to references \cite{Segura2015,Mahoney2009,Mller2019}.

The twist angle (before functionalization) and degree of functionalization are analyzed by Raman spectroscopy. All measurements were performed with a HORIBA LabRam HR Evo spectrometer with a laser wavelength \(\lambda\) = 532.17\,nm \((E_L\) = 2.33\,eV); laser focus <1 \textmu{}m and a 300 lines/mm grating. To avoid heating effects, laser powers below 0.5 mW were used. 
For lateral Raman maps a step size of 500\,nm was used.
The Raman spectra were calibrated by neon lines. 

The Quantum ESPRESSO suite was utilized to perform the DFT calculations \cite{QE-2009,QE-2017}. Structural relaxations were performed on \(6 \times 6 \), \(5 \times 5\), and \(7 \times 7\) bilayer graphene supercells which include one HATCN molecule. Norm conserving pseudopotentials from the SPMS library were used \cite{SPMS2023}. The PBE exchange-correlation functional \cite{pbe} was used in combination with the DFT-D3 vdW correction \cite{dft-d3} and the structures were optimized until the forces on each atom were less than 0.005\,eV/\AA. A converged plane wave energy cutoff of 60\,Ry (816.34\,eV) and a Monkhorst-Pack \(k\)-point grid with \(2 \times 2 \times 1\) \(k\) points were used. The dimensions of the unit cells perpendicular to the graphene layers were set to 35\,\AA. This was done to ensure that there is no interference between adjacent cells. However, due to the long ranging vdW interactions, significant stress in out-of-plane direction persists, which can be reduced by further increasing the size of the unit cell. We ensured the validity of our calculations by increasing the vertical size of the unit cells to 60\,\AA{} up to which only minimal geometric changes of the structures were observable, leaving the qualitative results unchanged.

To also investigate the interactions between neighboring HATCN molecules, charge density calculations were performed, in which the bilayer graphene supercells were doubled to \(12 \times 6 \), \(10 \times 5\), and \(14 \times 7\), respectively. The Monkhorst-Pack \(k\)-point grid was adjusted to \(1 \times 2 \times 1\).

For comparison with the self assembly of HATCN molecules, which was predicted on Ag(111) \cite{Mller2019}, a unit cell with two freestanding HATCN molecules was constructed. The plane wave energy cutoff and the Monkhorst-Pack \(k\)-point grid were adjusted to 70\,Ry (952.40\,eV) and \(3 \times 3 \times 1\) \(k\) points, respectively. The positions of the atoms were fixed within the \(xy\)-plane during the relaxation process.

\section{acknowledgement}

This work was financially supported by Deutsche Forschungsgemeinschaft (DFG, German Research Foundation) - 447264071 (INST 90/1183-1 FUGG) and 182849149 - SFB 953 -
Projects A1, A5, B4 \& B13.

The authors gratefully acknowledge the scientific support and HPC resources provided by the Erlangen National High Performance Computing Center (NHR@FAU) of the Friedrich-Alexander-Universität Erlangen-Nürnberg (FAU) under the NHR project b181dc. NHR funding is provided by federal and Bavarian state authorities. NHR@FAU hardware is partially funded by the German Research Foundation (DFG) – 440719683.

\printbibliography

@article{Kim2016,
  doi = {10.1021/acs.nanolett.5b05263},
  url = {https://doi.org/10.1021/acs.nanolett.5b05263},
  year = {2016},
  month = feb,
  publisher = {American Chemical Society ({ACS})},
  volume = {16},
  number = {3},
  pages = {1989--1995},
  author = {Kyounghwan Kim and Matthew Yankowitz and Babak Fallahazad and Sangwoo Kang and Hema C. P. Movva and Shengqiang Huang and Stefano Larentis and Chris M. Corbet and Takashi Taniguchi and Kenji Watanabe and Sanjay K. Banerjee and Brian J. LeRoy and Emanuel Tutuc},
  title = {van der Waals Heterostructures with High Accuracy Rotational Alignment},
  journal = {Nano Letters}
}

@article{Neumann2015,
  doi = {10.1038/ncomms9429},
  url = {https://doi.org/10.1038/ncomms9429},
  year = {2015},
  month = sep,
  publisher = {Springer Science and Business Media {LLC}},
  volume = {6},
  number = {1},
  author = {C. Neumann and S. Reichardt and P. Venezuela and M. Dr\"{o}geler and L. Banszerus and M. Schmitz and K. Watanabe and T. Taniguchi and F. Mauri and B. Beschoten and S. V. Rotkin and C. Stampfer},
  title = {Raman spectroscopy as probe of nanometre-scale strain variations in graphene},
  journal = {Nature Communications}
}

@article{Dean2010,
  doi = {10.1038/nnano.2010.172},
  url = {https://doi.org/10.1038/nnano.2010.172},
  year = {2010},
  month = aug,
  publisher = {Springer Science and Business Media {LLC}},
  volume = {5},
  number = {10},
  pages = {722--726},
  author = {C. R. Dean and A. F. Young and I. Meric and C. Lee and L. Wang and S. Sorgenfrei and K. Watanabe and T. Taniguchi and P. Kim and K. L. Shepard and J. Hone},
  title = {Boron nitride substrates for high-quality graphene electronics},
  journal = {Nature Nanotechnology}
}

@article{Mayorov2011,
  doi = {10.1021/nl200758b},
  url = {https://doi.org/10.1021/nl200758b},
  year = {2011},
  month = may,
  publisher = {American Chemical Society ({ACS})},
  volume = {11},
  number = {6},
  pages = {2396--2399},
  author = {Alexander S. Mayorov and Roman V. Gorbachev and Sergey V. Morozov and Liam Britnell and Rashid Jalil and Leonid A. Ponomarenko and Peter Blake and Kostya S. Novoselov and Kenji Watanabe and Takashi Taniguchi and A. K. Geim},
  title = {Micrometer-Scale Ballistic Transport in Encapsulated Graphene at Room Temperature},
  journal = {Nano Letters}
}

@article{Ye2014,
  doi = {10.1063/1.4866788},
  url = {https://doi.org/10.1063/1.4866788},
  year = {2014},
  month = mar,
  publisher = {{AIP} Publishing},
  volume = {104},
  number = {10},
  author = {Dong Ye and Shu-Qun Wu and Yao Yu and Lin Liu and Xin-Pei Lu and Yue Wu},
  title = {Patterned graphene functionalization via mask-free scanning of micro-plasma jet under ambient condition},
  journal = {Applied Physics Letters}
}

@article{Bao2021,
  doi = {10.1002/chem.202100941},
  url = {https://doi.org/10.1002/chem.202100941},
  year = {2021},
  month = may,
  publisher = {Wiley},
  volume = {27},
  number = {34},
  pages = {8709--8713},
  author = {Lipiao Bao and Baolin Zhao and Mhamed Assebban and Marcus Halik and Frank Hauke and Andreas Hirsch},
  title = {Covalent 2D Patterning,  Local Electronic Structure and Polarization Switching of Graphene at the Nanometer Level},
  journal = {Chemistry {\textendash} A European Journal}
}

@article{Hernndez2013,
  doi = {10.1016/j.carbon.2013.03.059},
  url = {https://doi.org/10.1016/j.carbon.2013.03.059},
  year = {2013},
  month = aug,
  publisher = {Elsevier {BV}},
  volume = {60},
  pages = {84--93},
  author = {Sandra C. Hern{\'{a}}ndez and Francisco J. Bezares and Jeremy T. Robinson and Joshua D. Caldwell and Scott G. Walton},
  title = {Controlling the local chemical reactivity of graphene through spatial functionalization},
  journal = {Carbon}
}

@article{Mohiuddin2009,
  title = {Uniaxial strain in graphene by Raman spectroscopy: $G$ peak splitting, Gr\"uneisen parameters, and sample orientation},
  author = {Mohiuddin, T. M. G. and Lombardo, A. and Nair, R. R. and Bonetti, A. and Savini, G. and Jalil, R. and Bonini, N. and Basko, D. M. and Galiotis, C. and Marzari, N. and Novoselov, K. S. and Geim, A. K. and Ferrari, A. C.},
  journal = {Phys. Rev. B},
  volume = {79},
  issue = {20},
  pages = {205433},
  numpages = {8},
  year = {2009},
  month = {May},
  publisher = {American Physical Society},
  doi = {10.1103/PhysRevB.79.205433},
  url = {https://link.aps.org/doi/10.1103/PhysRevB.79.205433}
}

@article{Carozo2011,
  doi = {10.1021/nl201370m},
  url = {https://doi.org/10.1021/nl201370m},
  year = {2011},
  month = oct,
  publisher = {American Chemical Society ({ACS})},
  volume = {11},
  number = {11},
  pages = {4527--4534},
  author = {Victor Carozo and Clara M. Almeida and Erlon H. M. Ferreira and Luiz Gustavo Can{\c{c}}ado and Carlos Alberto Achete and Ado Jorio},
  title = {Raman Signature of Graphene Superlattices},
  journal = {Nano Letters}
}

@article{Schpers2022,
  doi = {10.1088/2053-1583/ac7e59},
  url = {https://doi.org/10.1088/2053-1583/ac7e59},
  year = {2022},
  month = jul,
  publisher = {{IOP} Publishing},
  volume = {9},
  number = {4},
  pages = {045009},
  author = {A Sch\"{a}pers and J Sonntag and L Valerius and B Pestka and J Strasdas and K Watanabe and T Taniguchi and L Wirtz and M Morgenstern and B Beschoten and R J Dolleman and C Stampfer},
  title = {Raman imaging of twist angle variations in twisted bilayer graphene at intermediate angles},
  journal = {2D Materials}
}

@article{Dierke2022,
  doi = {10.1021/acsanm.1c04559},
  url = {https://doi.org/10.1021/acsanm.1c04559},
  year = {2022},
  month = mar,
  publisher = {American Chemical Society ({ACS})},
  volume = {5},
  number = {4},
  pages = {4966--4971},
  author = {Tobias Dierke and Daniela Dasler and Tamara Nagel and Frank Hauke and Andreas Hirsch and Janina Maultzsch},
  title = {Spatial Control of Graphene Functionalization by Patterning a 2D Substrate: Implications for Graphene Based van-der-Waals Heterostructures},
  journal = {{ACS} Applied Nano Materials}
}

@article{Edelthalhammer2020,
  doi = {10.1002/anie.202006874},
  url = {https://doi.org/10.1002/anie.202006874},
  year = {2020},
  month = oct,
  publisher = {Wiley},
  volume = {59},
  number = {51},
  pages = {23329--23334},
  author = {Konstantin Felix Edelthalhammer and Daniela Dasler and Lisa Jurkiewicz and Tamara Nagel and Sabrin Al-Fogra and Frank Hauke and Andreas Hirsch},
  title = {Covalent 2D-Engineering of Graphene by Spatially Resolved Laser Writing/Reading/Erasing},
  journal = {Angewandte Chemie International Edition}
}

@article{Cao2018,
  doi = {10.1038/nature26154},
  url = {https://doi.org/10.1038/nature26154},
  year = {2018},
  month = mar,
  publisher = {Springer Science and Business Media {LLC}},
  volume = {556},
  number = {7699},
  pages = {80--84},
  author = {Yuan Cao and Valla Fatemi and Ahmet Demir and Shiang Fang and Spencer L. Tomarken and Jason Y. Luo and Javier D. Sanchez-Yamagishi and Kenji Watanabe and Takashi Taniguchi and Efthimios Kaxiras and Ray C. Ashoori and Pablo Jarillo-Herrero},
  title = {Correlated insulator behaviour at half-filling in magic-angle graphene superlattices},
  journal = {Nature}
}

@article{Xie2020,
  doi = {10.1103/physrevlett.124.097601},
  url = {https://doi.org/10.1103/physrevlett.124.097601},
  year = {2020},
  month = mar,
  publisher = {American Physical Society ({APS})},
  volume = {124},
  issue = {9},
  pages = {097601},
  numpages = {6},
  author = {Ming Xie and A.{\hspace{0.167em}}H. MacDonald},
  title = {Nature of the Correlated Insulator States in Twisted Bilayer Graphene},
  journal = {Physical Review Letters}
}

@article{Lu2019,
  doi = {10.1038/s41586-019-1695-0},
  url = {https://doi.org/10.1038/s41586-019-1695-0},
  year = {2019},
  month = oct,
  publisher = {Springer Science and Business Media {LLC}},
  volume = {574},
  number = {7780},
  pages = {653--657},
  author = {Xiaobo Lu and Petr Stepanov and Wei Yang and Ming Xie and Mohammed Ali Aamir and Ipsita Das and Carles Urgell and Kenji Watanabe and Takashi Taniguchi and Guangyu Zhang and Adrian Bachtold and Allan H. MacDonald and Dmitri K. Efetov},
  title = {Superconductors,  orbital magnets and correlated states in magic-angle bilayer graphene},
  journal = {Nature}
}

@article{Yankowitz2019,
  doi = {10.1126/science.aav1910},
  url = {https://doi.org/10.1126/science.aav1910},
  year = {2019},
  month = mar,
  publisher = {American Association for the Advancement of Science ({AAAS})},
  volume = {363},
  number = {6431},
  pages = {1059--1064},
  author = {Matthew Yankowitz and Shaowen Chen and Hryhoriy Polshyn and Yuxuan Zhang and K. Watanabe and T. Taniguchi and David Graf and Andrea F. Young and Cory R. Dean},
  title = {Tuning superconductivity in twisted bilayer graphene},
  journal = {Science}
}

@article{Yoo2019,
  doi = {10.1038/s41563-019-0346-z},
  url = {https://doi.org/10.1038/s41563-019-0346-z},
  year = {2019},
  month = apr,
  publisher = {Springer Science and Business Media {LLC}},
  volume = {18},
  number = {5},
  pages = {448--453},
  author = {Hyobin Yoo and Rebecca Engelke and Stephen Carr and Shiang Fang and Kuan Zhang and Paul Cazeaux and Suk Hyun Sung and Robert Hovden and Adam W. Tsen and Takashi Taniguchi and Kenji Watanabe and Gyu-Chul Yi and Miyoung Kim and Mitchell Luskin and Ellad B. Tadmor and Efthimios Kaxiras and Philip Kim},
  title = {Atomic and electronic reconstruction at the van der Waals interface in twisted bilayer graphene},
  journal = {Nature Materials}
}

@article{Bistritzer2011,
  doi = {10.1073/pnas.1108174108},
  url = {https://doi.org/10.1073/pnas.1108174108},
  year = {2011},
  month = jul,
  publisher = {Proceedings of the National Academy of Sciences},
  volume = {108},
  number = {30},
  pages = {12233--12237},
  author = {Rafi Bistritzer and Allan H. MacDonald},
  title = {Moir{\'{e}} bands in twisted double-layer graphene},
  journal = {Proceedings of the National Academy of Sciences}
}

@article{Hidalgo2019,
  doi = {10.1021/acs.jpcc.9b02577},
  url = {https://doi.org/10.1021/acs.jpcc.9b02577},
  year = {2019},
  month = jun,
  publisher = {American Chemical Society ({ACS})},
  volume = {123},
  number = {24},
  pages = {15273--15283},
  author = {Francisco Hidalgo and Alberto Rubio-Ponce and Cecilia Noguez},
  title = {Tuning Adsorption of Methylamine and Methanethiol on Twisted-Bilayer Graphene},
  journal = {The Journal of Physical Chemistry C}
}

@article{Hidalgo2021,
author = {Hidalgo, Francisco and Rubio-Ponce, Alberto and Noguez, Cecilia},
title = {Cysteine Adsorption on Twisted-Bilayer Graphene},
journal = {The Journal of Physical Chemistry C},
volume = {125},
number = {49},
pages = {27314-27322},
year = {2021},
doi = {10.1021/acs.jpcc.1c08649},
}

@article{Majumdar2023,
  title = {Does twist angle affect the properties of water confined inside twisted bilayer graphene?},
  volume = {158},
  ISSN = {1089-7690},
  url = {http://dx.doi.org/10.1063/5.0139256},
  DOI = {10.1063/5.0139256},
  number = {3},
  pages = {034501},
  journal = {The Journal of Chemical Physics},
  publisher = {AIP Publishing},
  author = {Majumdar,  Jeet and Dasgupta,  Subhadeep and Mandal,  Soham and Moid,  Mohd and Jain,  Manish and Maiti,  Prabal K.},
  year = {2023},
  month = jan 
}

@article{Araki2022,
author = {Araki, Yuji and Solis-Fernandez, Pablo and Lin, Yung-Chang and Motoyama, Amane and Kawahara, Kenji and Maruyama, Mina and Gao, Yanlin and Matsumoto, Rika and Suenaga, Kazu and Okada, Susumu and Ago, Hiroki},
title = {Twist Angle-Dependent Molecular Intercalation and Sheet Resistance in Bilayer Graphene},
journal = {ACS Nano},
volume = {16},
number = {9},
pages = {14075-14085},
year = {2022},
doi = {10.1021/acsnano.2c03997},
}

@article{Schfer2016,
  doi = {10.1002/anie.201607427},
  url = {https://doi.org/10.1002/anie.201607427},
  year = {2016},
  month = oct,
  publisher = {Wiley},
  volume = {55},
  number = {47},
  pages = {14858--14862},
  author = {Ricarda A. Sch\"{a}fer and Konstantin Weber and Maria Kole{\'{s}}nik-Gray and Frank Hauke and Vojislav Krsti{\'{c}} and Bernd Meyer and Andreas Hirsch},
  title = {Substrate-Modulated Reductive Graphene Functionalization},
  journal = {Angewandte Chemie International Edition}
}

@article{Tahara2018,
  doi = {10.1021/acsnano.8b06681},
  url = {https://doi.org/10.1021/acsnano.8b06681},
  year = {2018},
  month = nov,
  publisher = {American Chemical Society ({ACS})},
  volume = {12},
  number = {11},
  pages = {11520--11528},
  author = {Kazukuni Tahara and Toru Ishikawa and Brandon E. Hirsch and Yuki Kubo and Anton Brown and Samuel Eyley and Lakshya Daukiya and Wim Thielemans and Zhi Li and Peter Walke and Shingo Hirose and Shingo Hashimoto and Steven De Feyter and Yoshito Tobe},
  title = {Self-Assembled Monolayers as Templates for Linearly Nanopatterned Covalent Chemical Functionalization of Graphite and Graphene Surfaces},
  journal = {{ACS} Nano}
}

@article{delaRie2022,
  doi = {10.1021/acs.jpcc.1c10266},
  url = {https://doi.org/10.1021/acs.jpcc.1c10266},
  year = {2022},
  month = jun,
  publisher = {American Chemical Society ({ACS})},
  volume = {126},
  number = {23},
  pages = {9855--9861},
  author = {Joris de la Rie and Mihaela Enache and Qiankun Wang and Wenbo Lu and Milan Kivala and Meike St\"{o}hr},
  title = {Self-Assembly of a Triphenylene-Based Electron Donor Molecule on Graphene: Structural and Electronic Properties},
  journal = {The Journal of Physical Chemistry C}
}

@article{Segura2015,
  doi = {10.1039/c5cs00181a},
  url = {https://doi.org/10.1039/c5cs00181a},
  year = {2015},
  publisher = {Royal Society of Chemistry ({RSC})},
  volume = {44},
  number = {19},
  pages = {6850--6885},
  author = {Jos{\'{e}} L. Segura and Rafael Ju{\'{a}}rez and Mar Ramos and Carlos Seoane},
  title = {Hexaazatriphenylene ({HAT}) derivatives: from synthesis to molecular design,  self-organization and device applications},
  journal = {Chemical Society Reviews}
}

@article{Mahoney2009,
  doi = {10.1039/b914290h},
  url = {https://doi.org/10.1039/b914290h},
  year = {2009},
  publisher = {Royal Society of Chemistry ({RSC})},
  volume = {19},
  number = {48},
  pages = {9221},
  author = {Stuart J. Mahoney and Mohamed M. Ahmida and Himadri Kayal and Nicholas Fox and Yo Shimizu and S. Holger Eichhorn},
  title = {Synthesis,  mesomorphism and electronic properties of nonaflate and cyano-substituted pentyloxy and 3-methylbutyloxy triphenylenes},
  journal = {Journal of Materials Chemistry}
}

@article{Christodoulou2014,
author = {Christodoulou, C. and Giannakopoulos, A. and Nardi, M. V. and Ligorio, G. and Oehzelt, M. and Chen, L. and Pasquali, L. and Timpel, M. and Giglia, A. and Nannarone, S. and Norman, P. and Linares, M. and Parvez, K. and M\"{u}llen, K. and Beljonne, D. and Koch, N.},
title = {Tuning the Work Function of Graphene-on-Quartz with a High Weight Molecular Acceptor},
journal = {The Journal of Physical Chemistry C},
volume = {118},
number = {9},
pages = {4784-4790},
year = {2014},
doi = {10.1021/jp4122408},
}

@article{Jeong2016,
author = {Jeong, Junkyeong and Park, Soohyung and Kang, Seong Jun and Lee, Hyunbok and Yi, Yeonjin},
title = {Impacts of Molecular Orientation on the Hole Injection Barrier Reduction: CuPc/HAT-CN/Graphene},
journal = {The Journal of Physical Chemistry C},
volume = {120},
number = {4},
pages = {2292-2298},
year = {2016},
doi = {10.1021/acs.jpcc.5b11535},
}

@article{Oh2017,
title = {Energy level alignment at the interface of NPB/HAT-CN/graphene for flexible organic light-emitting diodes},
journal = {Chemical Physics Letters},
volume = {668},
pages = {64-68},
year = {2017},
issn = {0009-2614},
doi = {https://doi.org/10.1016/j.cplett.2016.12.007},
url = {https://www.sciencedirect.com/science/article/pii/S0009261416309563},
author = {Eonseok Oh and Soohyung Park and Junkyeong Jeong and Seong Jun Kang and Hyunbok Lee and Yeonjin Yi},
}

@article{Mller2019,
  doi = {10.1002/smll.201901741},
  url = {https://doi.org/10.1002/smll.201901741},
  year = {2019},
  month = jul,
  publisher = {Wiley},
  volume = {15},
  number = {33},
  pages = {1901741},
  author = {Kathrin M\"{u}ller and Nico Schmidt and Stefan Link and Ren{\'{e}} Riedel and Julian Bock and Walter Malone and Karima Lasri and Abdelkader Kara and Ulrich Starke and Milan Kivala and Meike St\"{o}hr},
  title = {Triphenylene-Derived Electron Acceptors and Donors on Ag(111): Formation of Intermolecular Charge-Transfer Complexes with Common Unoccupied Molecular States},
  journal = {Small}
}

@article{pbe,
  title = {Generalized Gradient Approximation Made Simple},
  author = {Perdew, John P. and Burke, Kieron and Ernzerhof, Matthias},
  journal = {Phys. Rev. Lett.},
  volume = {77},
  issue = {18},
  pages = {3865--3868},
  numpages = {0},
  year = {1996},
  month = {Oct},
  publisher = {American Physical Society},
  doi = {10.1103/PhysRevLett.77.3865},
}

@article{QE-2009,
doi = {10.1088/0953-8984/21/39/395502},
year = {2009},
month = {sep},
publisher = {},
volume = {21},
number = {39},
pages = {395502},
author = {Paolo Giannozzi and Stefano Baroni and Nicola Bonini and Matteo Calandra and Roberto Car and Carlo Cavazzoni and Davide Ceresoli and Guido L Chiarotti and Matteo Cococcioni and Ismaila Dabo and Andrea Dal Corso and Stefano de Gironcoli and Stefano Fabris and Guido Fratesi and Ralph Gebauer and Uwe Gerstmann and Christos Gougoussis and Anton Kokalj and Michele Lazzeri and Layla Martin-Samos and Nicola Marzari and Francesco Mauri and Riccardo Mazzarello and Stefano Paolini and Alfredo Pasquarello and Lorenzo Paulatto and Carlo Sbraccia and Sandro Scandolo and Gabriele Sclauzero and Ari P Seitsonen and Alexander Smogunov and Paolo Umari and Renata M Wentzcovitch},
title = {QUANTUM ESPRESSO: a modular and open-source software project for quantum
simulations of materials},
journal = {Journal of Physics: Condensed Matter},
}

@article{QE-2017,
  title = {Advanced capabilities for materials modelling with Quantum ESPRESSO},
  volume = {29},
  ISSN = {1361-648X},
  url = {http://dx.doi.org/10.1088/1361-648X/aa8f79},
  DOI = {10.1088/1361-648x/aa8f79},
  number = {46},
  journal = {Journal of Physics: Condensed Matter},
  publisher = {IOP Publishing},
  author = {P Giannozzi and O Andreussi and T Brumme and O Bunau and M Buongiorno Nardelli and M Calandra and R Car and C Cavazzoni and D Ceresoli and M Cococcioni and N Colonna and I Carnimeo and A Dal Corso and S de Gironcoli and P Delugas and R A DiStasio and A Ferretti and A Floris and G Fratesi and G Fugallo and R Gebauer and U Gerstmann and F Giustino and T Gorni and J Jia and M Kawamura and H-Y Ko and A Kokalj and E Kucukbenli and M Lazzeri and M Marsili and N Marzari and F Mauri and N L Nguyen and H-V Nguyen and A Otero-de-la-Roza and L Paulatto and S Ponce and D Rocca and R Sabatini and B Santra and M Schlipf and A P Seitsonen and A Smogunov and I Timrov and T Thonhauser and P Umari and N Vast and X Wu and S Baroni },
  year = {2017},
  month = {oct},
  pages = {465901}
}

@article{SPMS2023,
title = {Soft and transferable pseudopotentials from multi-objective optimization},
journal = {Comput. Phys. Commun.},
volume = {283},
pages = {108594},
year = {2023},
issn = {0010-4655},
doi = {https://doi.org/10.1016/j.cpc.2022.108594},
author = {Mostafa Faghih Shojaei and John E. Pask and Andrew J. Medford and Phanish Suryanarayana},
}

@article{dft-d3,
    author = {Grimme, Stefan and Antony, Jens and Ehrlich, Stephan and Krieg, Helge},
    title = {A consistent and accurate ab initio parametrization of density functional dispersion correction (DFT-D) for the 94 elements H-Pu},
    journal = {The Journal of Chemical Physics},
    volume = {132},
    number = {15},
    pages = {154104},
    year = {2010},
    month = {04},
    issn = {0021-9606},
    doi = {10.1063/1.3382344},
    url = {https://doi.org/10.1063/1.3382344},
}

\end{document}